\documentclass[11pt]{article}
\usepackage{amssymb}
\newcommand{\ket}[1]{\left | #1 \right \rangle}
\newcommand{\bra}[1]{\left \langle #1 \right |}
\newcommand{\ave}[1]{\langle #1 \rangle}

\def\openone{\leavevmode\hbox{\small1\kern-3.8pt\normalsize1}}

\newcommand{\inner}[2]{ \langle #1 | #2 \rangle}

\newcommand{\beq}{\begin{equation}}
\newcommand{\eeq}{\end{equation}}
\newcommand{\beqa}{\begin{eqnarray}}
\newcommand{\eeqa}{\end{eqnarray}}

\begin{document}
\begin{center}
{\LARGE\bf Complex weak values in quantum measurement }\\
\bigskip
{\normalsize Richard Jozsa}\\
\bigskip
{\small\it Department of Computer Science, University of
Bristol,\\ Merchant Venturers Building, Bristol BS8 1UB U.K.} \\[4mm]
\date{today}
\end{center}

\begin{abstract} In the weak measurement formalism of Y. Aharonov et
al. the so-called weak value $A_w$ of any observable $A$ is
generally a complex number. We derive a physical interpretation of
its value in terms of the shift in the measurement pointer's mean
position and mean momentum. In particular we show that the mean
position shift contains a term jointly proportional to the
imaginary part of the weak value and the rate at which the pointer
is spreading
in space as it enters the measurement interaction.\\[2mm] PACS
numbers: 03.67.-a, 02.20.Qs
\end{abstract}
\bigskip
\section{Introduction}\label{intro}

In quantum mechanics the essential connection between theory and
experimental outcomes may be thought of as being embodied in the
formula
\begin{equation}\label{avval}
\ave{A} =  \bra{\psi_i} A \ket{\psi_i}
\end{equation}
for the measured mean value of an observable $A$ upon (strong)
measurement of a quantum system prepared in state $\ket{\psi_i}$.
The formalism of weak measurement \cite{AAV} developed by Y.
Aharonov and co-workers (c.f. \cite{BOOK} chapters 16,17 for a
review and further references therein) provides an alternative
foundation for quantum measurement theory. In this formalism the
above formula becomes replaced \cite{AAV} by a more general
expression:
\begin{equation}\label{wkval}
A_w = \frac{ \bra{\psi_f} A\ket{\psi_i}}{\inner{\psi_f}{\psi_i}}.
\end{equation}
$A_w$ is called the {\em weak value} of observable $A$ for a
quantum system pre-selected in state $\ket{\psi_i}$ and
post-selected in state $\ket{\psi_f}$ and it characterises the
observed outcomes of weak measurements. If $\ket{\psi_i}$ is an
eigenstate of $A$ then $\ave{A}$ and $A_w$ agree (both equalling
the corresponding eigenvalue) but more generally $A_w$ need not
lie within the range of eigenvalues and may even be complex. Thus
its significance is more subtle than the straightforward
interpretation of $\ave{A}$ as a measured mean value. In this note
we establish a physical interpretation for $A_w$ in its most
general context.

The formalism of weak measurement has two ingredients that differ
from the usual approach that leads to eq. (\ref{avval}): firstly
in addition to preparation of quantum systems in a given initial
state we also impose {\em post-selection} into a given final
state; secondly we consider a scenario in which the measurement
interaction is suitably weak so that after measurement the system
state is left largely intact. As a framework for our main results
we begin by briefly reviewing the weak measurement formalism and
the origin of the expression $A_w$ in eq. (\ref{wkval}).

Consider a quantum system prepared in state $\ket{\psi_i}$ upon
which we wish to measure $A$. For measurement process we use the
standard von Neumann paradigm \cite{VONN} introducing a pointer in
initial state $\ket{\phi}$ with wavefunction $\phi (q)$, and
interaction hamiltonian
\begin{equation}\label{int} H_{\rm int}=g(t)Ap\hspace{5mm} g(t)=g\delta
(t-t_0) \end{equation} where $g$ is a coupling constant and $p$ is
the pointer momentum conjugate to the position co-ordinate $q$.
Here we have taken the interaction to be impulsive at time $t=t_0$
(and the expression $Ap$ is shorthand for $(A\otimes I)(I\otimes
p)$, where the first and second slots refer to the system and
pointer respectively.)

After interaction the system and pointer are in joint state
$e^{-igAp}\ket{\psi_i}\ket{\phi}$ and we post-select the system on
state $\ket{\psi_f}$ resulting in the (sub-normalised) pointer
state
\begin{equation}\label{alpha} \ket{\alpha} =
\bra{\psi_f}e^{-igAp}\ket{\psi_i}\ket{\phi}. \end{equation} (Here
and hereafter we adopt units making $\hbar = 1$). In practice the
post-selection is achieved by running the process many times with
initial state $\ket{\psi_i}$ and after all tasks are completed we
perform a further final measurement of the projector $\Pi_f$ onto
$\ket{\psi_f}$ in each run. Then for statistical analysis of
measurement outcomes or any other considerations, we retain only
those runs for which $\Pi_f$ yielded 1. (The sub-normalisation of
$\ket{\alpha}$ reflects the probability of success in this $\Pi_f$
measurement). It is a remarkable fact that quantum theory allows
both pre- and post-selection of systems whereas classical physics
allows imposition of only either initial or final boundary
conditions, but not both (cf \cite{BOOK} \S 16.3).

It is a standard tenet of quantum theory that measurement
irrevocably disturbs a quantum system. The measurement interaction
eq. (\ref{int}) is said to be strong if the translated
wavefunctions $\phi (q-ga_i)$ for eigenvalues $a_i$ of $A$,
correspond to states that have negligible overlap. In that case
after the measurement interaction the pointer position will be
observed, on average, to have shifted by $g\ave{A}$. In contrast
to this standard scenario, the second basic ingredient in the
formalism of weak measurement is the requirement that the
measurement interaction eq. (\ref{int}) be suitably weak so that
we may obtain information about $A$ while the system state is left
largely intact. To restrict the strength of interaction we
consider the limit of small $g$, retaining only terms to first
order in $g$. Alternatively weakness may be imposed by requiring
$p$ to remain small which, by the $\Delta p\Delta q$ uncertainty
relation, corresponds to a limit of increasingly broad initial
wavefunctions of the pointer in the $q$ representation. In the
following we will work only with the limit of small $g$. In both
cases the translates $\phi (q-ga_i)$ will retain a large overlap
(of size $1-O(g)$ or $1-O(p)$). Expanding eq. (\ref{alpha}) to
terms of $O(g)$ yields:
\begin{eqnarray}
\ket{\alpha} & \approx & \bra{\psi_f} I-igAp\ket{\psi_i}\ket{\phi}
\nonumber \\
  & = & \inner{\psi_f}{\psi_i}\,\, \left( I-i g A_w p \right) \ket{\phi}
\label{ogexp2} \\
  & \approx &   \inner{\psi_f}{\psi_i} \,\, e^{-igA_wp} \ket{\phi}.
\label{ogexp3} \end{eqnarray} Thus it is clear that all subsequent
measurement properties of the pointer depend on the ingredients
$A$, $\ket{\psi_i}$ and $\ket{\psi_f}$ only through the single
c-number $A_w$.

From eq. (\ref{wkval}) we see that $A_w$ can generally be a
complex number and its effect on mean values of pointer variables,
such as the mean position and mean momentum, is not immediately
clear from eqs. (\ref{ogexp2},\ref{ogexp3}). Mathematically eq.
(\ref{ogexp3}) simply represents a translation $\phi (q-gA_w)$ of
the wavefunction by $gA_w$. However as the latter is generally
complex and we use the resulting translated function only along
the real $q$ axis, its quantum mean properties are now not simply
characterisable in terms of translates of those of $\phi (q)$. In
the literature only some special restricted cases have been
considered. Introduce the initial and final pointer means:
\begin{equation}\label{means}
\ave{q}_i=\bra{\phi} q\ket{\phi} \hspace{1cm} \ave{q}_f=
\frac{\bra{\alpha}q\ket{\alpha}}{\inner{\alpha}{\alpha}}
\end{equation}
and similarly the momentum means $\ave{p}_i$ and $\ave{p}_f$ with
$p$ replacing $q$ in the above. Also introduce the variances of
position and of momentum in the initial pointer state:
\begin{equation}\label{vars}
Var_q = \bra{\phi}q^2\ket{\phi} - \bra{\phi}q\ket{\phi}^2
\hspace{1cm} Var_p = \bra{\phi}p^2\ket{\phi} -
\bra{\phi}p\ket{\phi}^2. \end{equation} Then the following cases
have been noted \cite{BOOK,MJP}. (i) if $A_w$ is {\em real} then
$\ave{q}_f=\ave{q}_i+gA_w$; (ii) if $A_w$ is complex but the
pointer wavefunction $\phi(q)$ is {\em real-valued} then
$\ave{q}_f=\ave{q}_i + g Re(A_w)$ and $\ave{p}_f =
\ave{p}_i+2gIm(A_w) Var_p$; (iii) it has also been noted
(\cite{BOOK} p.237) that in the expression eq. (\ref{ogexp3}), the
imaginary part of $A_w$ contributes a non-unitary operation which
can thus be thought of as increasing or decreasing the size
$\inner{\alpha}{\alpha}$ of the pre- and post-selected ensemble of
runs.

\section{Complex weak values}

We now consider the most general case of complex $A_w$ and
complex-valued wavefunction $\phi(q)$. Our resulting general
formulae will display a novel role for the imaginary part of $A_w$
in the shift of pointer mean position. We will demonstrate the
following.

\noindent {\bf Theorem.} Let $A_w=a+ib$. Then after a weak von
Neumann measurement interaction on a system with pre- and
post-selected states $\ket{\psi_i}$ and $\ket{\psi_f}$, the mean
pointer position and momentum satisfy
\begin{eqnarray}
\ave{q}_f & = & \ave{q}_i+ga +gb\, (m\frac{d}{dt}Var_q)
\label{qgen}\\
\ave{p}_f & = & \ave{p}_i+2gb(Var_p) \label{pgen}
\end{eqnarray}
Here $m$ is the mass of the pointer and $\frac{d}{dt}Var_q$ is the
time derivative of its position variance as $t\rightarrow t_0$,
the time of the impulsive measurement interaction. $\Box$

Thus in particular there is a contribution to the pointer's mean
position shift that is proportional to the imaginary part of $A_w$
and the rate at which the pointer is spreading in space as it
enters the interaction.

To derive eq. (\ref{qgen}) we begin by substituting $p=-i\partial
/\partial q$ into eq. (\ref{ogexp2}). Retaining only terms to
$O(g)$ we get
\begin{equation}\label{aabar}
\alpha(q)\bar{\alpha}(q) = |\inner{\psi_f}{\psi_i}|^2 \left[
\phi\bar{\phi}-ga (
\phi'\bar{\phi}+\bar{\phi}'\phi)-igb(\phi'\bar{\phi}-\bar{\phi}'\phi)\right]
\end{equation}
where $\phi'$ denotes the space derivative and $\bar{\phi}$
denotes the complex conjugate. The coefficient of $ga$ is the
space derivative of the probability density $\phi\bar{\phi}$
whereas the coefficient of $gb$ is recognised as the spatial part
of the conserved probability current for $\ket{\phi}$. To exploit
these features we introduce
\begin{equation}\label{rsvars} \phi=Re^{iS}\hspace{1cm} \rho=R^2.
\end{equation}
Then \[ \alpha\bar{\alpha}= |\inner{\psi_f}{\psi_i}|^2 \left[
\rho-ga\rho'+gb(2\rho S') \right] \] and a straightforward
calculation to $O(g)$ gives (writing $\mu=\ave{q}_i$):
\begin{equation}
\ave{q}_f =  \frac{\int \bar{\alpha}q\alpha}{
\int\bar{\alpha}\alpha}
  =  \mu-ga\int q\rho '+gb\int 2\rho S' (q-\mu)
 \end{equation}
 where the integration is over all space. Integration by parts
 gives:
 \begin{equation}\label{eqqf} \ave{q}_f=\ave{q}_i+ga-gb\int
 (q-\mu)^2(\rho S')' .\end{equation}
Now consider the Schr\"{o}dinger equation of the pointer up to
time $t_0$ of interaction:
\[ i\phi_t=-\frac{1}{2m} \phi''+V(q)\phi .\]
Substituting eq. (\ref{rsvars}) and taking the imaginary part of
the resulting equation gives the continuity equation for
probability conservation: \begin{equation}\label{continuity}
\rho_t+(\rho\frac{S'}{m})'=0 . \end{equation} Hence $(\rho
S')'=-m\rho_t$ and eq. (\ref{eqqf}) finally gives
\begin{equation}\label{result}
\ave{q}_f=\ave{q}_i+ga+gb(m\frac{d}{dt}Var_q) \end{equation} as
claimed.

We note that a wavefunction is (instantaneously) real valued (up
to an overall constant phase) if and only if $S'=0$ and then $d\,
Var_q\, /dt$ is zero (via eq. (\ref{continuity}) giving
$\rho_t=0$) so we regain the previously quoted results (i) and
(ii) for the change in $\ave{q}$ in these restricted cases.

Next we present an alternative Heisenberg representation
derivation of eq. (\ref{result}) which generalises immediately to
other pointer observables (such as $p$) replacing $q$. Let $M$ be
any pointer observable. From eq. (\ref{ogexp2}) we get to $O(g)$:
\begin{eqnarray}
\ave{M}_f & = & \frac{\bra{\alpha}
M\ket{\alpha}}{\inner{\alpha}{\alpha}} \nonumber \\
 & = &
 \frac{\bra{\phi}M\ket{\phi}-igA_w\bra{\phi} Mp\ket{\phi}+ig\bar{A}_w
 \bra{\phi}
 pM\ket{\phi}}{\inner{\phi}{\phi}-igA_w\bra{\phi}p\ket{\phi}+ig\bar{A}_w
 \bra{\phi}p\ket{\phi}} \nonumber \\
  & = & \ave{M}_i+iga
  \ave{pM-Mp}_i+gb(\ave{pM+Mp}_i-2\ave{p}_i\ave{M}_i)\label{mexp}
  \end{eqnarray}
  (where for any observable $N$, $\ave{N}_i=\bra{\phi}N\ket{\phi}$
  is its mean value in state $\ket{\phi}$).

  For $M=q$ we have the commutation relations
  \[ [p,q]=-i \]
  and the Heisenberg equations of motion (with $H=p^2/2m+V(q)$):
  \begin{eqnarray}
     &i\frac{d}{dt}\ave{q}= \ave{[q,H]}=\frac{i\ave{p}}{m}  \nonumber
  \\  & i\frac{d}{dt}\ave{q^2}= \ave{[q^2,H]}=\frac{i\ave{pq+qp}}{m}.  \nonumber
  \end{eqnarray}
  Substitution of these into eq. (\ref{mexp}) immediately
  gives eq. (\ref{result}).

  If instead we set $M=p$ then $pM-Mp$ in the coefficient of $ga$ in
  eq. (\ref{mexp}) becomes zero and the coefficient of
  $gb$ becomes
  $2\ave{p^2}_i-2\ave{p}_i=2\, Var_p$ giving
  \[ \ave{p}_f=\ave{p}_i+2gb\, Var_p \]
  as claimed in the theorem. Note that the pointer observable $p$
  commutes with the measurement interaction hamiltonian $gAp$ so
  this shift in $\ave{p}$ is an artefact of post-selection rather
  than a quantum dynamical effect, in contrast to the more interesting
  case of the shift in $\ave{q}$.

\bigskip
\noindent {\large\bf Acknowledgements}\\ Thanks to G. Mitchison
for providing the Heisenberg representation derivation for the
change in $\ave{q}$. This work was presented at the workshop
``Weak Values and Weak Measurements: A New Approach to Reality in
Quantum Theory'', held at Arizona State University, June 2007
under the auspices of the Center for Fundamental Concepts in
Science, directed by P. Davies. It is a pleasure to acknowledge
discussions with participants, especially Y. Aharonov, A. Botero,
L. Diosi, G. Mitchison, S. Popescu and A. Steinberg amongst
others, and also to acknowledge financial support from the Center
as well as from the EPSRC QIPIRC network and the EC networks QAP
and QICS.

\end{document}